\documentclass[nofootinbib,pra,twocolumn,superscriptaddress]{revtex4-2}
\usepackage{float,graphicx,multirow, amsmath,color,dsfont}
\usepackage{amsmath,bm}
\usepackage{palatino}
\usepackage[colorlinks]{hyperref}
\usepackage[noabbrev]{cleveref}
\usepackage{enumerate}
\usepackage{amssymb,mathrsfs,dsfont}
\usepackage{amssymb,mathbbol}
\usepackage{amsfonts}
\usepackage{mathrsfs}
\usepackage{float}
\usepackage{xcolor,hyperref}
\definecolor{darkblue}{rgb}{0,0.0.1,0.3}
\definecolor{darkred}{rgb}{0.6,0.1,0}
\hypersetup{colorlinks,breaklinks,
	linkcolor=darkred,urlcolor=darkblue,
	anchorcolor=darkred,citecolor=
	darkred,pdfauthor=est, pdftitle= Estimation}
\usepackage{tikz}
\usepackage{mathbbol}
\usepackage{float}
\usepackage{array}
\usepackage{accents}

\newcommand{\vardbtilde}[1]{\widetilde{\raisebox{0pt}[0.95\height]{$\widetilde{#1}$}}}
\usepackage{braket}

\usepackage[normalem]{ulem} % For bibliography 
\newcommand{\etal}{\textit{et al}.\,}

\begin{document}	
\title{Optimal non-Gaussian operations  in difference-intensity detection and parity detection-based Mach-Zehnder interferometer}
\author{Manali Verma}
\email{manali.verma.31@gmail.com}
\affiliation{Department of Physical Sciences, Indian Institute of Science Education and Research Kolkata, Mohanpur-741246, West Bengal, India}
\author{Chandan Kumar}
\email{chandan.quantum@gmail.com}
\affiliation{Optics and Quantum Information Group, The Institute of Mathematical Sciences, CIT Campus, Taramani, Chennai 600113, India.}
\affiliation{Homi Bhabha National Institute, Training School Complex, Anushakti Nagar, Mumbai 400085, India.}
\author{Karunesh K. Mishra}
\email{karunesh.mishra@eli-np.ro}
\affiliation{Extreme Light Infrastructure-Nuclear Physics (ELI-NP), “Horia Hulubei” National R$\&$D Institute for Physics and Nuclear Engineering (IFIN-HH), 30 Reactorului Street, 077125 M˘agurele, jud. Ilfov, Romania}	
\author{Prasanta K. Panigrahi}
\email{pprasanta@iiserkol.ac.in}
\affiliation{Department of Physical Sciences, Indian Institute of Science Education and Research Kolkata, Mohanpur-741246, West Bengal, India}	
\begin{abstract}
  We investigate the benefits of probabilistic non-Gaussian   operations in phase estimation using difference-intensity and parity detection-based Mach-Zehnder interferometers (MZI).  We consider an experimentally implementable model to perform three different non-Gaussian operations, namely photon subtraction (PS), photon addition (PA), and photon catalysis (PC) on a single-mode squeezed vacuum (SSV) state. In difference-intensity detection-based MZI,  two PC operation is found to be the most optimal, while for parity detection-based MZI, two PA operation emerges as the most optimal process. We have also provided the corresponding squeezing and transmissivity parameters yielding best performance, making our study relevant for experimentalists. Further, we have derived the general expression  of moment-generating function, which shall be useful in exploring other detection schemes such as homodyne detection and quadratic homodyne detection.	
\end{abstract}
\maketitle
%%%%%%%%%%%%%%%%%%%%%%%%%%%%%%%%%%
\section{Introduction}

In the realm of optical instrumentation, the Mach-Zehnder interferometer (MZI) is a frequently utilized instrument for phase sensitivity measurement~\cite{Caves1981,Dowling-cp-2008,Giovannetti2011}. When exploiting the classical resources, the sensitivity of the MZI is limited by shot noise limit (SNL)~\cite{Caves1981}, whereas numerous non-classical states, including squeezed states, NOON states and Fock states are employed to surpass the SNL and approach the Heisenberg limit~\cite{Dowling,Giovannetti-science-2004,Hofmann-pra-2007, Anisimov-prl-2010,caves-prl-2013,Jeong-prl-2019}. This facilitates highly precise phase sensitivity measurements that are advantageous for various disciplines, including gravitational wave detection \cite{PhysRevLett.123.231107}, quantum-enhanced dark matter searches~\cite{Backes2021}, and biological  samples measurements~\cite{TAYLOR20161}.

In particular, single mode squeezed vacuum (SSV) state together with the single mode coherent state can be used as inputs to MZI for the estimation of unknown phase~\cite{Gard2017,caves-prl-2013,Seshadreesan2011,Ataman2018,mishra2022}. Squeezed states of light were initially observed by Slusher \etal in 1985~\cite{PhysRevLett.55.2409}, since then it has played a crucial role in quantum metrology for enhancing the phase sensitivity. The generation of highly squeezed states of light can be quite challenging~\cite{15dB}, therefore, we strive to find out ways to improve the phase sensitivity  even with  modest level of squeezing.

 Non-Gaussian operation is one such technique which has been employed in different quantum protocols such as squeezing and entanglement distillation~\cite{catalysisprl2002,catalysisnature2015,Swain:22,Multistep,distil}, quantum teleportation~\cite{tel2000,dellanno-2007,tel2009,catalysis15,catalysis17,wang2015,tele-2023,noisytele}, quantum key distribution~\cite{qkd-pra-2013,Ma-pra-2018, qk2019,zubairy-pra-2020}, and quantum metrology~\cite{gerryc-pra-2012,josab-2012,braun-pra-2014,josab-2016,pra-catalysis-2021,ill2008,ill2013,metro-thermal-arxiv, chandan-pra-23}, to enhance the performance.  Two of the most important examples of non-Gaussian operations are photon subtraction (PS) and photon addition (PA). In most of the theoretical studies, PS and PA operations are implemented via the annihilation and creation operators, $\hat{a}$ and $\hat{a}^\dagger$, respectively. However, annihilation and creation operators are non-unitary and cannot be directly implemented in the laboratory. 

  To circumvent this issue, we consider a realistic model for the implementation of non-Gaussian operations on a SSV state, combining it with  multiphoton Fock state $\ket{k}$ via a beam splitter of variable transmissivity, $\tau$, followed by the detection of $l$ photons [shown in Fig.~\ref{mzi}(b)]. The generated states are termed as photon-subtracted SSV (PSSSV) state,  photon-added SSV (PASSV) state, and  photon-catalyzed SSV (PCSSV) state, corresponding to three different non-Gaussian operations namely, PS, PA and photon catalysis (PC), which we collectively label as non-Gaussian SSV (NGSSV) states. These non-Gaussian operations have been experimentally realized in laboratory~\cite{Bellini,PhysRevLett.92.153601,doi:10.1126/science.1146204,doi:10.1126/science.1122858}.

In a recent paper~\cite{PhysRevA.101.063810}, it was demonstrated that the PS operation implemented via the annihilation operator $\hat{a}$ in difference intensity detection-based MZI, can enhance the phase sensitivity. Here, we extend the analysis to three different non-Gaussian operations, namely PS, PA, and PC,  using our proposed experimentally implementable model in difference intensity detection-based MZI and observe that not only PS but PC can also enhance the phase sensitivity, whereas PA fails to do so. It is worth noting that the results of Ref.~\cite{PhysRevA.101.063810}  can be obtained in the unit transmissivity limit of our realistic PS operation. In our work, we optimize the phase sensitivity over the transmissivity of the beam splitter involved in the implementation of the non-Gaussian operations. Since the beam splitter is considered to be an inexpensive device in quantum optical context~\cite{Wolf,Ivan}, one can select the beam splitter with required transmissivity $\tau$  for the implementation of various non-Gaussian operation.

We observe that for two PS (2-PS) operation, the phase sensitivity is not always maximized in the unit transmissivity limit. Therefore, the analysis of phase sensitivity presented in  our work, which incorporates the non-Gaussian operations based on an experimental model, offers the best insights into the maximum achievable phase sensitivity.

Given the probabilistic nature of the non-Gaussian operations, it is essential to take into account their success probability. Favourably, our experimental model enables us to consider the success probability of the non-Gaussian operations. Upon consideration of success probability, we find that 2-PC turns out to be the most optimal non-Gaussian operation for difference intensity detection-based MZI. We also provide optimal squeezing and transmissivity parameters, delivering the best performance, rendering our research valuable for experimentalists as well.

Subsequently, we consider the parity detection scheme and observe that all the three non-Gaussian operations (except 1-PC) can yield enhancement in phase sensitivity.  In comparison to the earlier work~\cite{chandan-pra-23}, where 1-PC yields an advantage, no such advantage is found in the present analysis when working at the phase matching condition of the coherent plus SSV state.  Further, upon consideration of success probability, 2-PA turns out to be the most optimal non-Gaussian operation.

This article is structured as follows. In Sec.~\ref{sec:mzi}, we briefly introduce the experimental setup of MZI explaining various non-Gaussian operations on SSV state. We then evaluate the phase sensitivity for difference-intensity detection-based MZI and carry out the analysis to find out the optimal non-Gaussian operation in Sec.~\ref{DID}. In Sec.~\ref{sec:pa}, we analyze the phase sensitivity for the parity-detection-based MZI to pinpoint the most optimal non-Gaussian operation. Finally, we outline our main results and provide directions for future research in  Sec.~\ref{sec:conc}.
%%%%%%%%%%%%%%%%%%%%%%%%%%%%%%%%%%
\section{Non-Gaussian operations  in MZI }\label{sec:mzi}
 \begin{figure}[h!]  
 	\begin{center}
 		\includegraphics[width=0.35\textwidth]{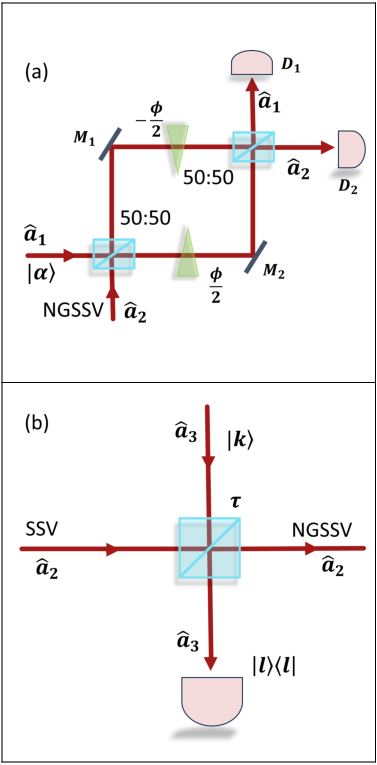}
 		\caption{(a) Schematic diagram of the MZI with coherent plus NGSSV state as inputs. (b) Schematic diagram for the implementation of different non-Gaussian operations on SSV state.
 		}
 		\label{mzi}
 	\end{center}
 \end{figure}
We consider a configuration for the balanced MZI, which consists of two input ports, two 50:50 beam splitters, two phase shifters and two output ports as shown in Fig.~\ref{mzi}(a). We introduce coherent state $\ket{\alpha}$ from one of the input port and NGSSV state from the other input port. The NGSSV state is achieved by performing different non-Gaussian operations on the SSV states, the schematic of which is given in Fig.~\ref{mzi}(b). The
 unknown phase introduced by the two phase shifters is determined using difference intensity detection and parity detection schemes. 

 \subsection{Implementation of non-Gaussian operation on SSV state}

We first describe the implementation of non-Gaussian operations on SSV state as depicted in Fig.~\ref{mzi}(b).  
The mode associated with the SSV state is represented by the annihilation and creation operators $\hat{a}_2$ and $\hat{a}_2^\dagger$. Similarly, the auxiliary mode corresponding to   Fock state $\ket{k}$ is represented by the annihilation and creation operators $\hat{a}_3$ and $\hat{a}_3^\dagger$. The corresponding phase space variables for these modes are $\xi_2=(q_2,p_2)^T$, and $\xi_3=(q_3,p_3)^T$.
Since the SSV state is a zero centered Gaussian state, it can be specified by the   covariance matrix,
\begin{equation}
V=\frac{1}{2}\left(\begin{array}{cc}
e^{-2 r} & 0 \\
0 & e^{2 r}
\end{array}\right),
\end{equation}
where $r$ is the squeezing parameter.
Consequently, the Wigner  function of the SSV state can be written as~\cite{weedbrook-rmp-2012}
\begin{equation}
  \label{wig:ssv}\mathcal{W}_{\hat{a}_2}(\xi_2 )   = \pi^{-1} \exp(-e^{2r}q_2^2-e^{-2r}p_2^2).
\end{equation}
Similarly, the Wigner   function of the Fock state $|k\rangle$  turns out to be
	\begin{equation}\label{wig:fock}
		\mathcal{W}_{\hat{a}_3}(\xi_3 )=\frac{(-1)^k}{\pi}\exp  \left( 
		-q_3^2-p_3^2 \right)\,\mathcal{L}_{k}\left[ 2(q_3^2+p_3^2) \right],
	\end{equation}
with $\mathcal{L}_k\{\bullet\}$ being the Laguerre polynomial of $k$th order.
 The joint Wigner function of the SSV and the Fock state is simply the product of the individual Wigner distribution functions:
	\begin{equation}
		\mathcal{W}_{ \hat{a}_2, \hat{a}_3}(\xi_{23}) =  \mathcal{W}_{\hat{a}_2}(\xi_2 ) \mathcal{W}_{\hat{a}_3}(\xi_3 ),
	\end{equation}
 where $\xi_{23}=(q_2,p_2,q_3,p_3)^T$. For the implementation of non-Gaussian operations, we combine the SSV state with Fock state $|k\rangle$ via a beam splitter of variable transmissivity $\tau$. The beam splitter entangles the two modes and  the corresponding Wigner function can be written as
  \begin{equation}
		\widetilde{\mathcal{W}}_{ \hat{a}_2, \hat{a}_3}(\xi_{23} ) = 		\mathcal{W}_{ \hat{a}_2, \hat{a}_3}(B_{\hat{a}_2, \hat{a}_3}(\tau)^{-1} \xi_{23}),
	\end{equation} 
 where $B_{\hat{a}_2, \hat{a}_3}(\tau)$ denotes  the action of the beam splitter on the phase space variables  $\xi_{23}$ 
 \begin{equation}\label{beamsplitter}
		B_{\hat{a}_2, \hat{a}_3}(\tau) = \begin{pmatrix}
			\sqrt{\tau} \,\mathbb{1}_2& \sqrt{1-\tau} \,\mathbb{1}_2 \\
			-\sqrt{1-\tau} \,\mathbb{1}_2&\sqrt{\tau} \,\mathbb{1}_2
		\end{pmatrix},
	\end{equation}
   with $\mathbb{1}_2$ being the second order identity matrix.
A photon number resolving detector is employed to detect photons on the output auxiliary mode, $\hat{a_3}$. Whenever the detector registers  $l$ photons, it heralds the generation of NGSSV state on the output mode $\hat{a_2}$. The Wigner function  of the NGSSV state is given by 
\begin{equation}\label{detect}
		\begin{aligned}
		\widetilde{\mathcal{W}}_{ \hat{a}_2 }(\xi_{2} )= 2 \pi\int  d^2 \xi_3  
		\widetilde{ \mathcal{ W}}_{\hat{a}_2, \hat{a}_3}(\xi_{23} ) 
			\times 
			 \widetilde{\mathcal{W}}_{\hat{a}_3}(\xi_3 ) .\\
		\end{aligned}
	\end{equation}
 where $\widetilde{\mathcal{W}}_{\hat{a}_3}(\xi_3 )$ corresponds to the Wigner function of the Fock state $\ket{l}$. Equation~(\ref{detect})
integrates out to\footnote{We note some typos in the corresponding equation of Ref.~\cite{chandan-pra-23}.} 
\begin{equation}\label{eq4}
	\widetilde{\mathcal{W}}_{ \hat{a}_2 }(\xi_{2} ) = a_0 \bm{\widehat{F}} \exp \left( \xi_2^T A_1 \xi_2 + u^T A_2 \xi_2+ u^T A_3 u \right),
\end{equation}
where
\begin{equation}
\begin{aligned}
    a_0=&\sqrt{\frac{\lambda ^2-1}{\lambda ^2 \tau ^2-1}}, \quad \lambda= \tanh r.
\end{aligned}
    \end{equation}
Differential operator $\bm{\widehat{F}} $ is defined as 
\begin{equation}\label{F operator}
	\bm{\widehat{F}} = \frac{(-2)^{k+l}}{\pi \, k! \, l!} \frac{\partial^{k}}{\partial\,u_1^{k}} \frac{\partial^{k}}{\partial\,v_1^{k}} \frac{\partial^{l}}{\partial\,u_2^{l}} \frac{\partial^{l}}{\partial\,v_2^{l}}\{ \bullet \}_{\substack{u_1= v_1=0\\ u_2= v_2=0}},\\
\end{equation}
and the column vector $u$ is defined as $u=(u_1,v_1,u_2,v_2)^T$.
The explicit forms of the matrices $A_1,A_2$ and $A_3$
are provided in Eqs.
\labelcref{A1,A2,A3} of Appendix \ref{Appendix A}. 

The aforementioned Wigner function~(\ref{eq4}) is  not normalized. Thus, the  normalized Wigner function is given by
\begin{equation}\label{normalizedW}
     \vardbtilde{{\mathcal{W}}}_{ \hat{a}_2 }(\xi_{2} ) =  P^{-1} \widetilde{\mathcal{W}}_{ \hat{a}_2 }(\xi_{2} ),
 \end{equation}
  where $P$ is the success probability of non-Gaussian operations, which can be calculated as follows:
  
\begin{equation}\label{probeqq}
\begin{aligned}
    P &= \int d^2 \xi_2    \widetilde{{\mathcal{W}}}_{ \hat{a}_2 }(\xi_{2} )\\
 &= \pi a_0 \bm{\widehat{F}} \exp \left(   u^T A_4 u \right),
\end{aligned}
\end{equation}
where $A_4$ is given by the Eq.~(\ref{A4}) of Appendix~\ref{Appendix A}.
 One can implement different non-Gaussian operations on SSV state by suitably choosing the input multiphoton Fock state  and  detecting the desired number of photons in the detector. To be more specific, one can perform PS, PA or PC operation on SSV state under the condition $k<l$, $k>l$ or $k=l$, respectively. The states generated are called as PSSSV, PASSV, and PCSSV states, respectively. In this article, we have set $k=0$ for PS and $l=0$ for PA. These non-Gaussian operations convert the SSV state from Gaussian to non-Gaussian. 
   
The Wigner distribution function of the NGSSV~(\ref{normalizedW}) is quite general and from there one can obtain the Wigner distribution function of non-Gaussian states in different limits. For instance, the Wigner distribution function of the ideal PSSSV state can be obtained in the limit $\tau \rightarrow 1$ with  $k=0$. The ideal PSSSV state are represented by $ \mathcal{N}_s \hat{a}^{l} \ket{\text{SSV}}$, where $\mathcal{N}_s$ is the normalization factor.
	Similarly, 	the Wigner distribution function of the ideal PASSV state can be obtained in the limit $\tau \rightarrow 1$ with  $l=0$. 
		The ideal PASSV state are represented by $ \mathcal{N}_a \hat{a}{_2^{\dagger }}^{k}  \ket{\text{SSV}}$, where $\mathcal{N}_a$ is the normalization factor.

 \subsection{Coherent  and NGSSV states as input to MZI}
Now, the coherent state $\ket{\alpha}$ and the generated NGSSV states~(\ref{normalizedW}) are used as a resource to the MZI.   The mode corresponding to coherent state $|\alpha\rangle$ is represented by the annihilation and creation operators $\hat{a}_1$ and $\hat{a}_1^\dagger$.  The mode corresponding to NGSSV state is represented by the annihilation and creation operators $\hat{a}_2$ and $\hat{a}_2^\dagger$, as shown in Fig.~\ref{mzi}(a).
 The Wigner   function of the coherent state $| \alpha \rangle$  turns out to be 
\begin{equation}\label{coh}
    \mathcal{W}_{ \hat{a}_1 }(\xi_{1} )  = (\pi)^{-1}\exp \left[ -(q_1-d_x)^2-(p_1-d_p)^2\right],
\end{equation} 
where $\alpha=(d_x+id_p)/\sqrt{2}$.  Here $d_x$ and $d_p$ are small displacements along $q$ and $p$ quadratures. The Wigner function of the input state of the MZI is the product of the Wigner function of the coherent state and NGSSV state,
\begin{equation}
\mathcal{W}_{\text{in}}(\xi_{1,2} ) = \mathcal{W}_{ \hat{a}_1 }(\xi_{1} )  \vardbtilde{\mathcal{W}}_{ \hat{a}_2 }(\xi_{2} ). 
  \end{equation}
The Schwinger representation~\cite{schwinger2001angular} of the $\text{SU}(2)$ algebra~\cite{yurke-1986} is used as a tool to describe the collective action of the MZI. The generators of the $\text{SU}(2)$ algebra are typically denoted as $J_1,J_2$ and $J_3$. In the Schwinger representation, these generators can be expressed in terms of the annihilation and creation operators of the input modes as follows:  
\begin{equation}
	\begin{aligned}
		J_1 = &\frac{1}{2}(\hat{a}^\dagger_1\hat{a}_2+\hat{a}_1\hat{a}^\dagger_2),\\
		J_2 = &\frac{1}{2i}(\hat{a}^\dagger_1\hat{a}_2-\hat{a}_1\hat{a}^\dagger_2),\\
		J_3 = &\frac{1}{2}(\hat{a}^\dagger_1\hat{a}_1-\hat{a}^\dagger_2\hat{a}_2).
	\end{aligned}
\end{equation}
These operators satisfy
 the commutation relations of the $\text{SU}(2)$ algebra, $[J_i,J_j] = i \epsilon_{ijk}J_k $ and are
 also known as the angular momentum operator. The total action of the MZI can be represented as a product of unitary operators associated with the individual components of the interferometer. The unitary operators for the first  and the second  beam splitters are given by $e^{-i(\pi/2)\hat{J}_1}$ and $e^{i(\pi/2)\hat{J}_1}$, respectively. The combined action of the two phase shifters is represented by the unitary operator $e^{i \phi \hat{J}_3}$. Hence, the total action of the MZI is given by,
\begin{equation}
\label{UMZI}	U_{MZI}= e^{-i(\pi/2)J_1}e^{i \phi J_3}e^{i(\pi/2)J_1}=e^{-i \phi J_2}.
\end{equation}
which corresponds to the following transformation matrix, acting on the phase space variables $\xi_{12}=(q_1,p_1,q_2,p_2)^T$:
 \begin{equation} 
	S =  \begin{pmatrix}
		\cos (\phi/2) \,\mathbb{1_2}& -	\sin (\phi/2) \,\mathbb{1_2} \\
		\sin (\phi/2) \,\mathbb{1_2}& \cos (\phi/2) \,\mathbb{1_2}
	\end{pmatrix}.
\end{equation}
The action of the transformation matrix, on the Wigner function of the two input modes of the MZI is given by~\cite{arvind1995, weedbrook-rmp-2012}
\begin{equation}\label{mzioutput}
	\mathcal{W}_{\text{out}}(\xi_{1,2} )=\mathcal{W}_{\text{out}}(\xi_{1},\xi_{2}) =\mathcal{W}_{\text{in}}(S^{-1}\xi_{1,2} ).
\end{equation}
This represents the Wigner function of the output state.
%%%%%%%%%%%%%%%%%%%%%%%%%%%%%%%%%%
 \section{Evaluation of phase sensitivity}\label{sec:ph}

 Using the error propagation formula, the phase uncertainty  introduced by the phase shifters of the MZI for a specific measurement $\hat{O}$  can be written as  
\begin{equation}\label{phasesens}
	\Delta \phi = \frac{\sqrt{\langle\hat{O}^2\rangle -	\langle\hat{O}\rangle^2}}{|\partial  	\langle\hat{O}\rangle/\partial \phi|}.
\end{equation} Our aim is to find the input NGSSV state, which minimizes the phase uncertainty $\Delta \phi$, thereby maximizing the phase sensitivity. 

\subsection{Difference-intensity detection }\label{DID}
Coherent plus squeezed vacuum state  has been extensively used  as inputs in difference-intensity detection-based MZI \cite{Ataman2018,mishra2022}.
 The measurement operator  corresponding to difference intensity detection, $\hat{O}= \hat{a}_1^\dagger \hat{a}_1-\hat{a}_2^\dagger \hat{a}_2$, which is also called as photon number difference operator.   For the evaluation of phase uncertainty~(\ref{phasesens}), we need to compute the mean and variance of the  operator $\hat{O}$.   The operator $\hat{O}^2$ can be written as
\begin{equation}
\begin{aligned}
     \hat{O}^2=  (\hat{a}_1^\dagger \hat{a}_1-\hat{a}_2^\dagger \hat{a}_2)^2 = (\hat{a}_1^\dagger \hat{a}_1)^2 +(\hat{a}_2^\dagger \hat{a}_2)^2-2 \hat{a}_1^\dagger \hat{a}_1\hat{a}_2^\dagger \hat{a}_2.
\end{aligned}
  \end{equation}
   For calculation purpose we need to express $\hat{O}$ and $\hat{O}^2$ in symmetric ordered form. 
The operator $\hat{a}_i^\dagger \hat{a}_i =	\frac{1}{2}\left( \hat{q}_i^2+\hat{p}_i^2 -1 \right), $ appearing in $\hat{O}$ and $\hat{O}^2$ is already in symmetric ordered form. Similarly, we can also write $(\hat{a}_i^\dagger \hat{a}_i)^2$ in symmetric ordered form as follows:
\begin{equation}
 (\hat{a}_i^\dagger \hat{a}_i)^2=  \frac{1}{4} \bigg[ \hat{q}_i^4+\hat{p}_i^4
	-2\hat{q}_i^2-2\hat{p}_i^2 + 2 {}_{\bm{:}}^{\bm{:}}	\hat{q}_i^2\hat{p}_i^2 {}_{\bm{:}}^{\bm{:}} \bigg],
\end{equation}
where ${}_{\bm{:}}^{\bm{:}}  \bullet {}_{\bm{:}}^{\bm{:}} $ denotes symmetrically ordered operator.
    We compute the mean of symmetrically ordered operator using the Wigner function~(\ref{mzioutput})  as follows:
\begin{equation}
	\left\langle  {}_{\bm{:}}^{\bm{:}} \hat{q_1}^{n_1} \hat{p_1}^{m_1} \hat{q_2}^{n_2} \hat{p_2}^{m_2} {}_{\bm{:}}^{\bm{:}} \right\rangle = \int d^4 \xi\, q_1^{n_1} p_1^{m_1} q_2^{n_2} p_2^{m_2} \mathcal{W}_{\text{out}}(\xi_{1,2} ).
\end{equation}

 This integral  can be evaluated using the
parametric differentiation technique:
\begin{equation}\label{para}
	\mathcal{M}_{n_1,m_1}^{n_2,m_2}= \bm{\widehat{D} } \int d^4 \xi\,
	e^{  x_1 q_1+y_1 p_1+x_2 q_2+y_2p_2 }
	\mathcal{W}_{\text{out}}(\xi_{1,2} ) ,
\end{equation}
where
\begin{equation}\label{D}
	\bm{\widehat{D} } =   \frac{\partial^{n_1}}{\partial\, x_1^{n_1}} \frac{\partial^{m_1}}{\partial\, y_2^{m_1}}\frac{\partial^{n_2}}{\partial\, x_2^{n_2}} \frac{\partial^{m_2}}{\partial\, y_2^{m_2}} \{ \bullet \}_{x_1=y_1=x_2=y_2=0}.
\end{equation}
The integral~(\ref{para}) evaluates to 
\begin{equation}\label{mom}
\begin{aligned}
  \mathcal{M}_{n_1,m_1}^{n_2,m_2}= \frac{\pi g_0}{P} &\bm{\widehat{D} } \bm{\widehat{F}} \exp \big(   u^T G_1 u +u^T G_2 x \\
  &+ x^T G_3 x+ x^T G_4  \big) 
\end{aligned}
\end{equation}
where $g_0=\sqrt{ \left(\lambda ^2 \tau ^2-1\right)/\left(\lambda ^2-1\right)}$
and the column vector $x$ is defined as $x=(x_1,y_1,x_2,y_2)^T$. $\bm{\widehat{F}}$, 
$P$ and $\bm{\widehat{D} } $ are defined in Eq. (\ref{F operator}), Eq. (\ref{probeqq}) and Eq. (\ref{D}), respectively.
The explicit form of the matrices $G_1, G_2, G_3$ and $G_4$ are provided in Eqs. \labelcref{G1,G2,G3,G4} of Appendix \ref{Appendix B}.

 The quantity $\mathcal{M}_{n_1,m_1}^{n_2,m_2}$  is similar to the moment generating function and different moments of symmetrically ordered operators in the phase uncertainty expression can be evaluated using this.
By setting appropriate values of the parameters $n_1$, $m_1$, $n_2$, and $m_2$, we obtain analytical expression of phase uncertainty Eq.~(\ref{phasesens}) for different NGSSV states.

\begin{figure}[h!]  
\includegraphics[scale=1]{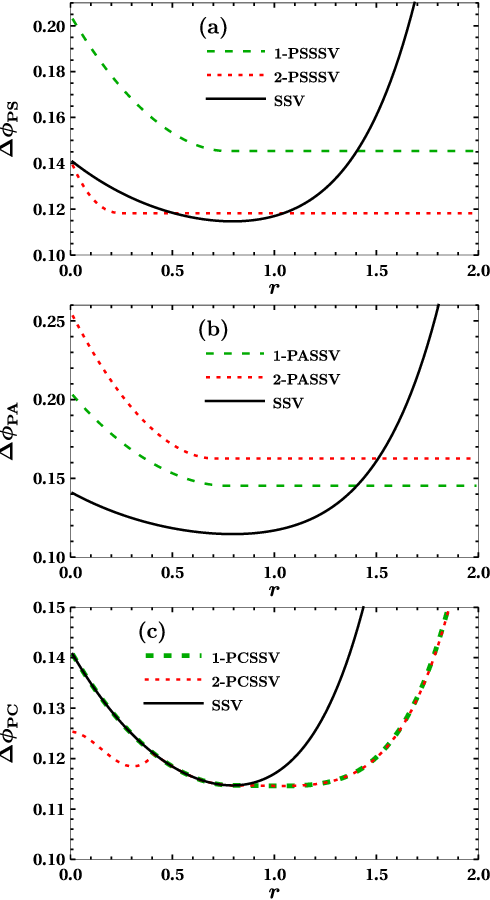}
\caption{Phase uncertainty $\Delta \phi$ as a function of squeezing parameter ($r$) for different NGSSV states in comparison with SSV state.}
\label{DI}
\end{figure}

Now we show the plot of $\Delta \phi$ optimized over the transmissivity $\tau$ of the beam splitter for different NGSSV states as a function
of squeezing parameter $r$ in Fig.~\ref{DI}. 
We have set $d_{x}=10$ and $d_{p}=0$, which corresponds to the phase matching condition of the coherent plus SSV state~\cite{Liu2013,Preda2019}. Further, for numerical analysis, we have set the phase $\phi=\pi/2$ rad. We use these numerical values throughout the analysis of difference-intensity detection based MZI. 

As can be seen from Fig.~\ref{DI}(a), 1-PSSSV state does not enhance the phase sensitivity, whereas the 2-PSSSV state enhances the phase sensitivity as compared to the SSV state for lower values of $r \in (0, 0.5)$. A similar trend was observed for ideal 1-PS and 2-PS operations in Ref.~\cite{PhysRevA.101.063810}. Similarly, one can see in Fig.~\ref{DI}(c) that the 2-PCSSV state shows better sensitivity as compared to the SSV state for small values of $r$. However, it is worth noticing that PA operations are inefficient in enhancing the phase sensitivity for smaller values of $r$, which can be seen from Fig.~\ref{DI}(b).

\begin{figure}[h!]  
\includegraphics[scale=1]{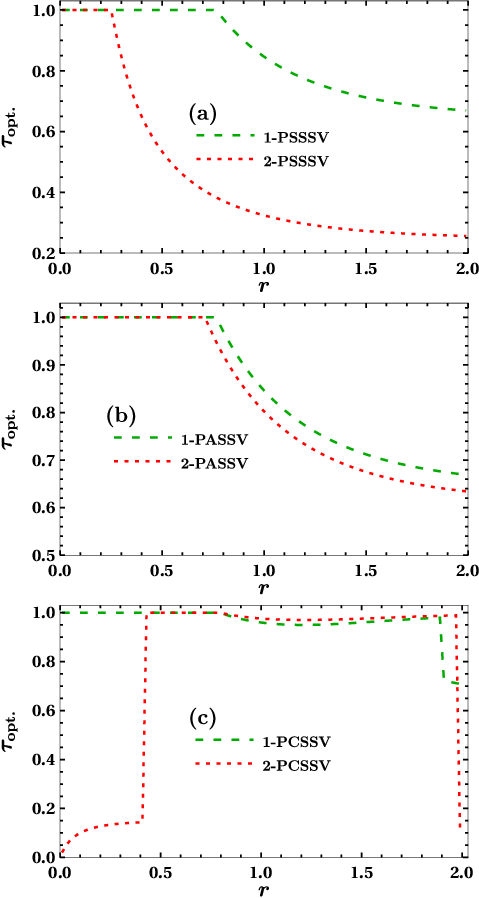}
\caption{Optimal beam splitter transmissivity $\tau$ (minimizing the phase uncertainty $\Delta \phi$ in Fig.~\ref{DI}) as a function of squeezing parameter ($r$) of the NGSSV state.
}
\label{OTVSDI}
\end{figure}

Figure~\ref{OTVSDI} depicts the optimal  transmissivity $\tau$   as a function of the
squeezing parameter $r$ corresponding to the minimization of $\Delta \phi$ in Fig.~\ref{DI}. 
We notice a surprising result in the case of 2-PS operation, where the phase uncertainty is not minimized in the unit transmissivity limit for the range of squeezing parameters, $r \in (0.25, 0.5)$ where 2-PS operation offers an advantage.  We stress that   our study resembles the ideal photon subtraction considered in Ref.~\cite{PhysRevA.101.063810} in the unit transmissivity limit.  Since the phase uncertainty is not always minimized in this limit, it highlights the importance of our analysis.  Similarly, one can see in  Fig.~\ref{OTVSDI}(c) that for the case of 2-PC operation, 
the phase uncertainty is not minimized in the  unit transmissivity limit.

\begin{figure}[h!] 
\includegraphics[scale=1]{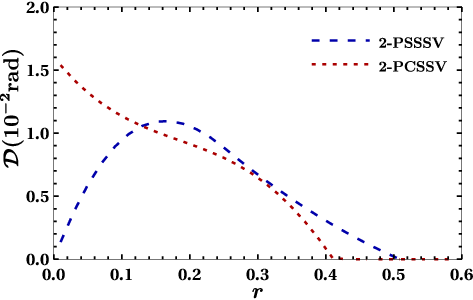}
\caption{Difference in phase uncertainty $\mathcal{D} = \Delta \phi_\text{SSV} - \Delta \phi_\text {NGSSV}$ as a function of squeezing parameter ($r$).}
\label{Relative_phase_sensitivity_vs_squeezing_parameter_inten_diff}
\end{figure}

To find out the squeezing parameter yielding the maximum enhancement in phase sensitivity and also compare the two phase sensitivity enhancing operations 2-PS and 2-PC,
we consider the difference of the phase uncertainty defined as follows: 
\begin{equation}\label{merit2}
    \mathcal{D}  = \Delta \phi_\text{SSV} - \Delta \phi_\text{NGSSV}.
\end{equation}
We plot the quantity $\mathcal{D}$ optimized over the transmissivity $\tau$ as a function of $r$ in Fig.~\ref{Relative_phase_sensitivity_vs_squeezing_parameter_inten_diff} for 2-PSSSV and 2-PCSSV states.  It is observed that for smaller values of the squeezing parameter, $r \in (0,0.12)$, 2-PC  is more advantageous operation as compared to 2-PS. However, for the remaining range of the squeezing parameter, 2-PS  becomes more advantageous.

Since $\Delta \phi_\text{SSV}$ is independent of transmissivity, the transmissivity maximizing the quantity $\mathcal{D}$ is same as the transmissivity minimizing  $\Delta \phi_\text{NGSSV}$. Consequently,  the optimal transmissivity maximizing the quantity $\mathcal{D}$ for 2-PSSSV and 2-PCSSV states turn out to be the one shown in Fig.~\ref{OTVSDI}(a) and (c), respectively.  

\subsubsection{Consideration of success probability}\label{success}
Since the considered non-Gaussian operations are probabilistic in nature, it is of paramount importance to take into account the success probability while discussing the advantages rendered by non-Gaussian states. To provide a relevant scenario, we consider the 2-PSSSV   and   2-PCSSV states and  analyze the corresponding difference in phase sensitivity $\mathcal{D}$  and probability as a  function of transmissivity ($\tau$). For 2-PSSSV state, we take the squeezing parameter $r=0.15$, where the $\mathcal{D}$ attains a maximum value (Fig.~\ref{Relative_phase_sensitivity_vs_squeezing_parameter_inten_diff}). Similarly, $\mathcal{D}$ for 2-PCSSV state is maximized in the zero squeezing limit. Therefore, to make the model experimentally implementable, we consider the squeezing parameter $r=0.05$.
\begin{figure}[h!] 
\includegraphics[scale=1]{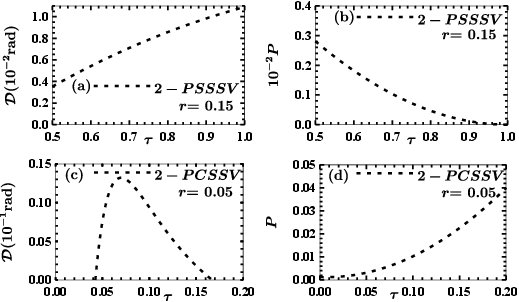}
\caption{ Left row shows difference in phase sensitivity $\mathcal{D}$ as a   function of transmissivity $\tau$. Right row shows success probability  as a function of transmissivity $\tau$.}
\label{prob_1d_diff_int}
\end{figure}

The plots for 2-PSSSV state in Figs.~\ref{prob_1d_diff_int}(a)-(b), shows that the difference in phase sensitivity $\mathcal{D}$ is maximized in the unit transmissivity limit; however, the corresponding success probability $(P)$ approaches zero. Such a scenario represents zero resource utilization and is impractical from an experimental point of view. To utilize the resource optimally, we adjust the transmissivity to trade off the enhancement in phase sensitivity against the success probability. To that end, we define the quantity 
\begin{equation}
    \mathcal{R} =\mathcal{D} \times P.
\end{equation}
Our aim is to maximize the product $ \mathcal{R}$ by adjusting the transmissvity $\tau$ for each value of squeezing parameter.
Similarly, for 2-PCSSV state [Figs.~\ref{prob_1d_diff_int}(c)-(d)], we can choose the appropriate transmissivity $\tau$ to maximize the product $ \mathcal{R}$.

\begin{figure}[h!]
\includegraphics[scale=1]{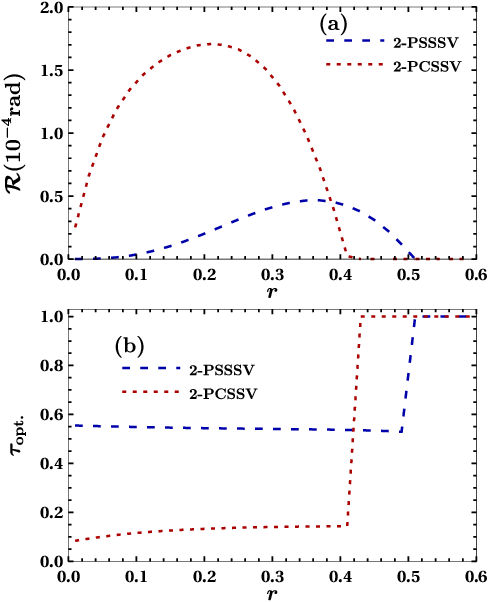}
\caption{ (a) Product of difference in phase sensitivity and success probability $\mathcal{R} = \mathcal{D} \times P$ as a function of squeezing parameter ($r$). (b) The corresponding optimal transmissivity of the beam splitter as a function of squeezing parameter ($r$).}
\label{Success_probability_and_tau_vs_squeezing_parameter_inten_diff}
\end{figure}

 We plot the quantity $ \mathcal{R}$ as a function of squeezing parameter ($r$),  in Fig.~\ref{Success_probability_and_tau_vs_squeezing_parameter_inten_diff}(a), which is optimized over the transmissivity $\tau$ of the beam splitter. We observe that 2-PC yields a better performance over 2-PS operation for low range of squeezing parameter.

We have also shown the optimal transmissivity corresponding to the maximization of $\mathcal{R}$ in Fig.~\ref{Success_probability_and_tau_vs_squeezing_parameter_inten_diff}(b), which shows that the transmissivity for 2-PC lies in the range, $\tau=0.1$.

We have summarized the maximum value of the product $\mathcal{R}$ and optimal parameters for 2-PS and 2-PC operations in the Table~\ref{table1}. These parameters can be used by experimentalists to obtain the best performance.

\begin{table}[hbt!]
		\caption{\label{table1}
			Maximum value of the product $\mathcal{R}$ and optimal parameters for  2-PS and 2-PC operations}
		\renewcommand{\arraystretch}{1.5}
		\begin{tabular}{ |p{1.5cm}|>{\centering\arraybackslash} p{1.5cm} |p{0.6cm}|p{0.6cm} |>{\centering\arraybackslash}p{1cm}|>{\centering\arraybackslash}p{1cm}|p{1cm}|}
			\hline \hline
			Operation & $10^4 \times \mathcal{R}_\text{max}$ (rad) &  $r_\text{opt}$ & $\tau_\text{opt}$& ${\Delta \phi}_\text{SSV}$ (rad) & $10^2\times \mathcal{D}$ (rad) & $10^2 \times P$\\  \hline \hline
   2-PS & 0.47 & 0.36 & 0.54 & 0.12 & 0.33 & 1.4 \\\hline
   2-PC & 1.70& 0.21 & 0.13 &0.13 &0.87 & 2.0\\ \hline \hline 
    	\end{tabular}
	\end{table}

 %------------------------------------------------------------------------------------------------ 
 \subsection{Parity detection}\label{sec:pa}
Coherent plus squeezed vacuum state  has been  used as inputs in parity detection-based MZI \cite{Seshadreesan2011}.
 We reexamine the work of Ref.~\cite{chandan-pra-23}, where the effect of non-Gaussian operations were explored in parity detection-based MZI. In the current study, we employ a coherent state with zero displacement along $\hat{p}$-quadrature\footnote{Here we have taken  $d_x=2$ and $d_p=0$ while Ref.~\cite{chandan-pra-23} considered $d_x=d_p=2$.  }, which corresponds to the phase matching condition of the coherent plus SSV state as input to the MZI~\cite{Seshadreesan2011}. The parity detection is performed on the mode $\hat{a}_2$ of Fig.~\ref{mzi}. 
The corresponding measurement operator is given by
\begin{equation}
	\hat{\Pi}_{\hat{a}_2} =  \exp\left( i \pi   \hat{a}^{\dagger}_2 \hat{a}_2 \right)=  (-1)^{\hat{a}_2^{\dagger}\hat{a}_2}.
\end{equation}
For the observable $\hat{O}=\hat{\Pi}_{\hat{a}_2}$, we can compute the phase uncertainty~(\ref{phasesens}) using the Wigner function approach.
  Since, $ \hat{\Pi} ^2  $  is the identity operator, $\langle  \hat{\Pi}^2  \rangle=1$. 
We can evaluate the average of the parity operator\footnote{A typo in the corresponding equation of Ref.~\cite{chandan-pra-23}: Missing probability in the denominator.} using the Wigner distribution function~(\ref{mzioutput}) as follows~\cite{Seshadreesan2011,Birrittella-2021,chandan-pra-23},
\begin{equation}\label{apari1}
\begin{aligned}
    \langle \hat{\Pi}_{\hat{a}_2} \rangle &  = \pi \int \, d^2\xi_1 \, \mathcal{W}_{\text{out}} (\xi_1,0),\\
    &= \frac{\pi p_1}{P} \bm{\widehat{F}} \exp \left(  d_x p_2+  u^T Q_1 u+ d_x Q_2 \right),
\end{aligned}
\end{equation}
where we have set $d_p=0$.  Here, \begin{equation}
    \begin{aligned}
   p_1=  \sqrt{\frac{\lambda ^2-1}{(\lambda  \tau  \cos (\phi ))^2-1}}, \,\,
    p_2=\frac{(\lambda  \tau +1) (\cos (\phi )-1)}{2-2 \lambda  \tau  \cos (\phi )}.
    \end{aligned}
\end{equation}
 $\bm{\widehat{F}}$ and P are defined in Eq.~(\ref{F operator})  and Eq.~(\ref{probeqq}), respectively.
The explicit form of the matrices $Q_1$ and $Q_2$ 
are provided in Eqs.~\labelcref{Q1,Q2} of the Appendix \ref{Appendix C}.

\begin{figure}[h!]
\includegraphics[scale=1]{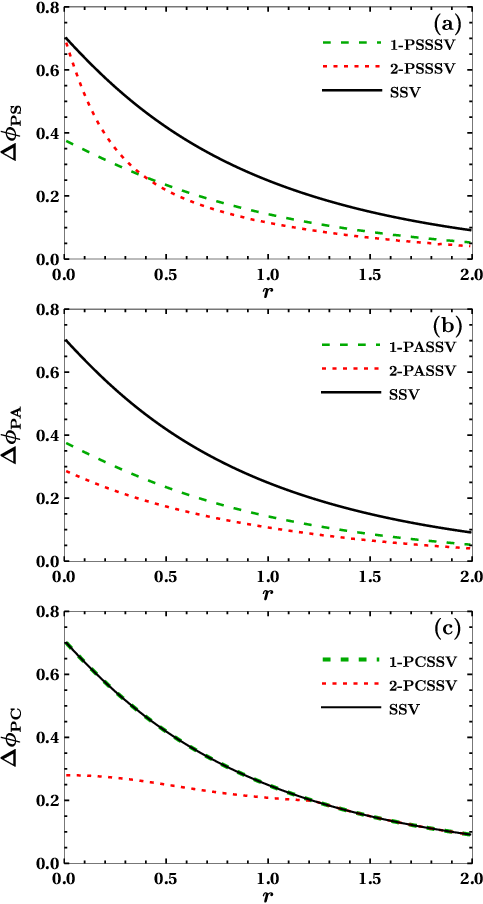}
\caption{Phase uncertainity $\Delta \phi$ as a function of squeezing parameter ($r$) for different NGSSV states in comparison with SSV state.}
\label{Fig:parity_vs_squeezing_parameter_v01}
\end{figure}

We optimize the phase uncertainty over the transmissivity $\tau$ for different values of the squeezing parameter $r$ and show the results in Fig.~\ref{Fig:parity_vs_squeezing_parameter_v01}. 
For the  analysis of parity detection based MZI we have
set $d_{x}=2$, $d_{p}=0$ and $\phi=0.01$\footnote{The phase matching condition of the coherent plus SSV
state for parity detection is $d_p=0$ and $\phi=0$~\cite{Seshadreesan2011}.} rad. 
We observe from Fig.~\ref{Fig:parity_vs_squeezing_parameter_v01}(a) that both 1-PSSSV and 2-PSSSV states enhance  the phase sensitivity as compared to the SSV state. Similarly, from Fig.~\ref{Fig:parity_vs_squeezing_parameter_v01}(b) one can see that 1-PASSV and 2-PASSV states also enhance the phase sensitivity. However, for PC operation [Fig.~\ref{Fig:parity_vs_squeezing_parameter_v01}(c)], the 1-PCSSV state fails to enhance the phase sensitivity, but the 2-PCSSV state enhances the phase sensitivity for smaller values of $r$. The advantage shown for the 1-PCSSV state in Ref.~\cite{chandan-pra-23} was an artifact of working at non-phase matching condition ($d_x=d_p =2$).
 
\begin{figure}[h!] 
\includegraphics[scale=1]{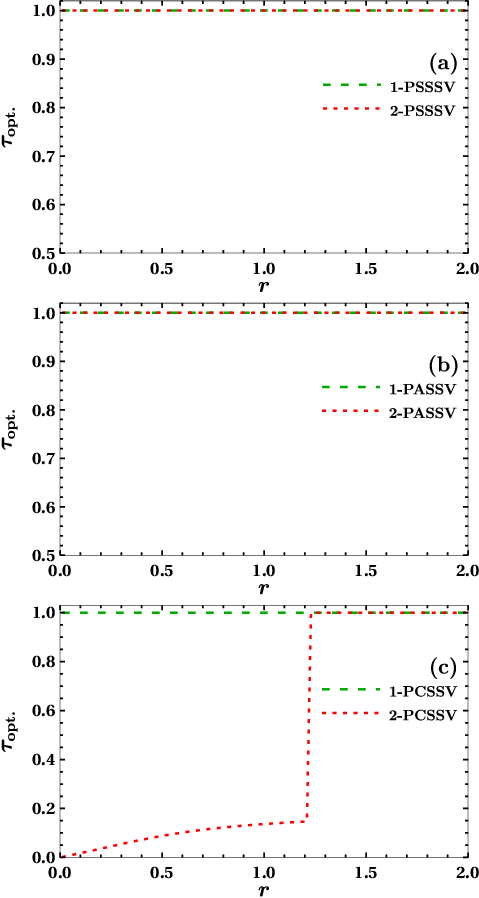}
\caption{Optimal beam splitter transmissivity $\tau$ (minimizing the phase uncertainty $\Delta \phi$ in Fig.~\ref{Fig:parity_vs_squeezing_parameter_v01}) as a function of squeezing parameter ($r$) of the NGSSV state.
}
\label{Fig:tau_vs_squeezing_parameter_v01_parity}
\end{figure}

Figure~\ref{Fig:tau_vs_squeezing_parameter_v01_parity} represents the optimal  transmissivity $\tau$ corresponding to the minimization of $\Delta \phi$ in Fig.~\ref{Fig:parity_vs_squeezing_parameter_v01}.  We notice that for PS and PA operation the phase uncertainity is minimized in the unit transmissivity limit, $\tau\to 1$. However, in the case of 2-PC operation, the phase uncertainty is minimized in the  non-unit transmissivity limit for the range of squeezing parameters, $r \in (0, 1.2)$ where 2-PC operation offers an advantage.

\begin{figure}[h!] 
\includegraphics[scale=1]{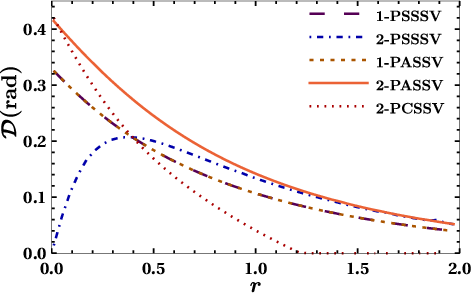}
\caption{Difference in phase uncertainty $\mathcal{D} = \Delta \phi_\text{SSV} - \Delta \phi_\text{NGSSV}$ as a function of squeezing parameter ($r$) for different NGSSV states.}
\label{Relative_phase_sensitivity_vs_squeezing_parameter_parity}
\end{figure}

Further, we plot the quantity $\mathcal{D}$ optimized over the transmissivity $\tau$ as a function of $r$ in Fig.~\ref{Relative_phase_sensitivity_vs_squeezing_parameter_parity} for PSSSV, PASSV and 2-PCSSV states.  It is observed that 2-PA is more advantageous operation for the complete range of squeezing parameter, ($r$).   

\subsubsection{Consideration of success probability}
We consider the 1-PSSSV, 1-PASSV and 2-PCSSV state and  analyze the corresponding difference in phase sensitivity $\mathcal{D}$  and success probability as a function of transmissivity ($\tau$) at $r=0.05$. 
\begin{figure}[h!] 
\includegraphics[scale=1]{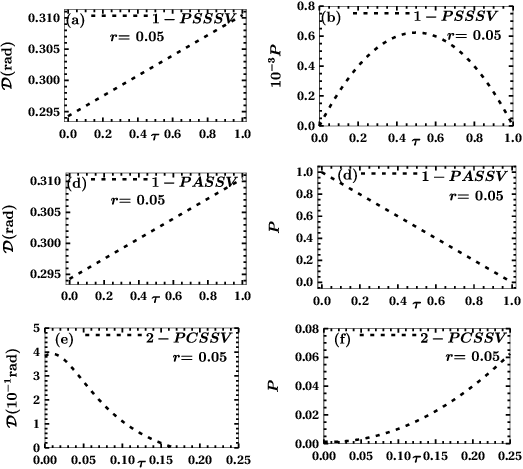}
\caption{ Left row shows difference in phase sensitivity $\mathcal{D}$ as a   function of transmissivity $\tau$. Right row shows success probability  as a function of transmissivity $\tau$.}
\label{prob_1d_parity}
\end{figure}
The results are shown in Fig.~\ref{prob_1d_parity}, which shows  that the difference in phase sensitivity $\mathcal{D}$ is maximized in the unit transmissivity limit for PSSSV and PASSV states; however, the corresponding success probability $(P) $ approaches zero. Similarly, for PCSSV state, the difference in phase sensitivity $\mathcal{D}$ is maximized in the zero transmissivity limit; however, the success probability approaches zero in this limit.  These scenarios are not feasible to implement experimentally. 

\begin{figure}[h!]
\includegraphics[scale=1]{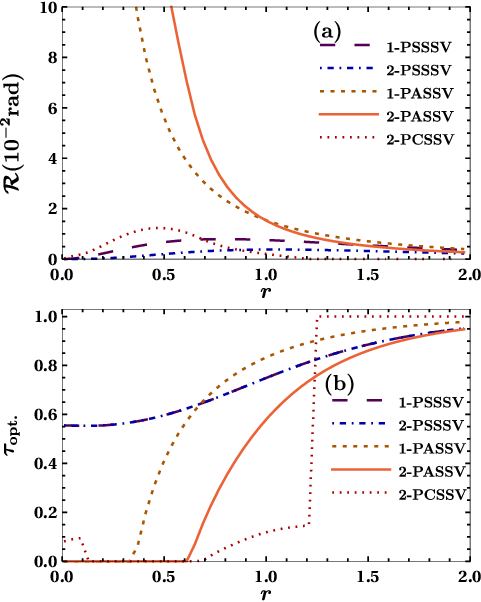}
\caption{  (a) Product of difference in phase sensitivity and success probability $\mathcal{R} = \mathcal{D} \times P$ as a function of squeezing parameter ($r$) for different NGSSV states.}
\label{Success_probability_and_tau_vs_squeezing_parameter_parity}
\end{figure}

To obtain an experimentally implementable  scenario, we calculate and maximize the quantity  $\mathcal{R}$ by adjusting $\tau$ as we did earlier in Sec.~\ref{success} for difference intensity detection case.  We plot the quantity  $\mathcal{R}$ optimized over the transmissivity $\tau$ as a function of $r$  in Fig.~\ref{Success_probability_and_tau_vs_squeezing_parameter_parity}(a). One can see that the maximum enhancement is obtained at $r=0$, similarly we have also shown the optimal transmissivity corresponding to the maximization of $\mathcal{R}$ in Fig.~\ref{Success_probability_and_tau_vs_squeezing_parameter_parity}(b), which shows that the transmissivity for 2-PA lies in the range, $\tau=0$, which cannot be achieved experimentally.  

We have summarized the maximum value of the product $\mathcal{R}$ and optimal parameters for PS, PA and 2-PC operations, in the Table~\ref{table2}.

The states for which optimal squeezing and transmissivity parameters approaches zero, can be experimentally implemented at low parameter regime such as $\tau=0.01$ and $r=0.01$. We have shown the corresponding values of product $\mathcal{R}$ and other parameters in Table~\ref{table3} for PA operation. We can see that the decrease is only by a minimal amount in the value of $\mathcal{R}$.

\begin{table}[H]
\centering
\caption{\label{table2}
Maximum value of the product $\mathcal{R}$ and optimal parameters for   non-Gaussian operations}
\renewcommand{\arraystretch}{1.5}
\begin{tabular}{ |p{1.5cm}|>{\centering\arraybackslash} p{1.5cm} |p{0.6cm}|p{0.6cm} |>{\centering\arraybackslash}p{1cm}|>{\centering\arraybackslash}p{1cm}|p{1cm}|}
\hline \hline
Operation & $10^{3} \times \mathcal{R}_\text{max}$ (rad)  &  $r_\text{opt}$ & $\tau_\text{opt}$& ${\Delta \phi}_\text{SSV}$ (rad)& $\mathcal{D}$ (rad)& $10^{2} \times P$\\ \hline \hline
 1-PS & 7.9 & 0.79 & 0.79 & 0.31 & 0.09 & 8.6 \\\hline
 2-PS & 3.7 & 1.04 & 0.76 &0.24 &0.07 & 5.4\\ \hline 
 1-PA & 329.1& 0 & 0 &0.71&0.33&   100\\ \hline  
 2-PA & 418.4 & 0 & 0 &0.71 &0.42 & 100\\ \hline 
 2-PC & 12.4& 0.47 & 0 &0.43 &0.14 & 8.7\\ \hline \hline 
 \end{tabular}
	\end{table}

 \begin{table}[H]
\centering
\caption{\label{table3}
Maximum value of the product $\mathcal{R}$  for PA operations at $r=0.01$ and $\tau=0.01$.}
\renewcommand{\arraystretch}{1.5}
\begin{tabular}
{ |p{1.5cm}|>{\centering\arraybackslash} p{1.5cm}|>{\centering\arraybackslash}p{1cm}|>{\centering\arraybackslash}p{1cm}|p{1cm}|}

\hline \hline
Operation & $10^{3} \times \mathcal{R}_\text{max}$ (rad)& ${\Delta \phi}_\text{SSV}$ (rad)& $\mathcal{D}$ (rad)& $10^{2} \times P$\\ \hline \hline
  
 1-PA & 318.9 &0.70&0.32& 99\\ \hline  
 2-PA & 403.2  &0.70 &0.41 & 98\\ \hline \hline

 \end{tabular}
	\end{table}

%-----------------------------------------------------------------------------------------

%%%%%%%%%%%%%%%%%%%%%%%%%%%%%%%%%%

\section{Conclusion}\label{sec:conc}

We investigated the advantages offered by different non-Gaussian operations in phase estimation using difference-intensity and parity-detection-based Mach-Zehnder Interferometers (MZI), with a coherent plus non-Gaussian single mode squeezed vacuum  (NGSSV) state as the two inputs. We considered  a realistic scheme for implementing three different non-Gaussian operations, namely, photon subtraction (PS), photon addition (PA), and
photon catalysis (PC), on the single mode squeezed vacuum (SSV) state. Our investigation
of the phase sensitivity highlighted that, PS and PC can significantly enhance the phase sensitivity  for difference-intensity detection scheme. Moreover, all the three non-Gaussian operations showed potential in enhancing phase sensitivity with the parity detection scheme. When the success probability is factored in the performance, two PCSSV state turns out to be the most optimal operation for difference intensity detection scheme and the two PASSV state showed superiority in the parity detection scheme.

We have also provided analytical expression for the moment generating function~(\ref{mom})  which will prove invaluable for investigating other detection schemes such as homodyne detection and quadratic homodyne detection~\cite{Jeong-pra-17}.  Another direction that we intend to pursue is to study the relative resilience of phase sensitivity of difference-intensity and parity detection scheme in lossy environment and whether unbalanced MZI can provide advantage over the balanced interferometer in non-Gaussian state based phase estimation~\cite{unbalanced-PRL,mishra2022}.

It is known that PS operation
can be used for designing quantum heat engines~\cite{Walmsley,Filip,Scarani,Zanin, Buller}, a systematic investigation of the same using other non-Gaussian operations like PA and PC is required. Similarly, one can also explore the role of non-Gaussian operations in inducing sub-Planck structure in phase space~\cite{Zurek2001,Utpal,Jitesh,Arman:21,Akhtar}.
 
 \section*{Acknowledgement}
	 M. Verma would like to thank IISER-Kolkata for the hospitality and financial support from the Interdisciplinary Cyber Physical Systems (ICPS) program of the Department of Science and Technology (DST), India through Grant No. DST/ICPS/QuEST/Theme-1/2019/6. 
  %KKM acknowledges the European Regional Development Fund and Competitiveness Operational Programme (1/07.07.2016, COP, ID Grant No. 1334) for financial support.....

	\appendix
 \begin{widetext}
\section{Matrices corresponding to Wigner distributions and success probability}\label{Appendix A}

Here, we provide the expressions of the matrices $A_1,A_2$ and $A_3$
which appears in the Wigner distribution of the NGSSV state~(\ref{eq4}):

    \begin{equation}\label{A1}
   A_1=\left(
\begin{array}{cc}
 \dfrac{2}{\lambda  \tau -1}+1 & 0 \\
 0 & 1-\dfrac{2}{\lambda  \tau +1} \\
\end{array}
\right),
    \end{equation}

    \begin{equation}\label{A2}
        A_2=\frac{1}{4 \left(\lambda ^2 \tau ^2-1\right)}\left(
\begin{array}{cccc}
 -\lambda  (\tau -1) \tau  & \tau -1 & -\lambda  (\tau -1) \sqrt{\tau } & \sqrt{\tau } \left(1-\lambda ^2 \tau \right) \\
 \tau -1 & -\lambda  (\tau -1) \tau  & \sqrt{\tau } \left(1-\lambda ^2 \tau \right) & -\lambda  (\tau -1) \sqrt{\tau } \\
 -\lambda  (\tau -1) \sqrt{\tau } & \sqrt{\tau } \left(1-\lambda ^2 \tau \right) & \lambda -\lambda  \tau  & \lambda ^2 (\tau -1) \tau  \\
 \sqrt{\tau } \left(1-\lambda ^2 \tau \right) & -\lambda  (\tau -1) \sqrt{\tau } & \lambda ^2 (\tau -1) \tau  & \lambda -\lambda  \tau  \\
\end{array}
\right),
    \end{equation}

    \begin{equation}\label{A3}
      A_3=\frac{\sqrt{1-\tau}}{\lambda ^2 \tau ^2-1}  
 \left(
\begin{array}{cc}
 -\lambda  \tau -1 & i (\lambda  \tau -1) \\
 \lambda  \tau +1 & i (\lambda  \tau -1) \\
 -\lambda  \sqrt{\tau } (\lambda  \tau +1) & -i \lambda  \sqrt{\tau } (\lambda  \tau -1) \\
 \lambda  \sqrt{\tau } (\lambda  \tau +1) & -i \lambda  \sqrt{\tau } (\lambda  \tau -1) \\
\end{array}
 \right).
    \end{equation}
Here, $A_4$ is the matrix corresponding to the success probability (\ref{probeqq}) of the non-Gaussian operations,
     \begin{equation}\label{A4}
     A_4=\frac{1}{4 \left(\lambda ^2 \tau ^2-1\right)}  \left(
\begin{array}{cccc}
 -\lambda  (\tau -1) \tau  & 1-\tau  & \lambda  (\tau -1) \sqrt{\tau } & \sqrt{\tau } \left(1-\lambda ^2 \tau \right) \\
 1-\tau  & -\lambda  (\tau -1) \tau  & \sqrt{\tau } \left(1-\lambda ^2 \tau \right) & \lambda  (\tau -1) \sqrt{\tau } \\
 \lambda  (\tau -1) \sqrt{\tau } & \sqrt{\tau } \left(1-\lambda ^2 \tau \right) & \lambda -\lambda  \tau  & -\lambda ^2 (\tau -1) \tau  \\
 \sqrt{\tau } \left(1-\lambda ^2 \tau \right) & \lambda  (\tau -1) \sqrt{\tau } & -\lambda ^2 (\tau -1) \tau  & \lambda -\lambda  \tau  \\
\end{array}
\right).
     \end{equation}
 
\section{Matrices corresponding to Moment Generating Function}\label{Appendix B}
The matrices, $G_1,G_2,G_3$ and $G_4$ corresponds to the moment generating function (\ref{mom}) of the main text,
 
    \begin{equation}\label{G1}
     G_1=\frac{1}{4 \left(1-\lambda ^2 \tau ^2\right)}   \left(
\begin{array}{cccc}
 \lambda  (\tau -1) \tau  & \tau -1 & -\lambda  (\tau -1) \sqrt{\tau } & \sqrt{\tau } \left(\lambda ^2 \tau -1\right) \\
 \tau -1 & \lambda  (\tau -1) \tau  & \sqrt{\tau } \left(\lambda ^2 \tau -1\right) & -\lambda  (\tau -1) \sqrt{\tau } \\
 -\lambda  (\tau -1) \sqrt{\tau } & \sqrt{\tau } \left(\lambda ^2 \tau -1\right) & \lambda  (\tau -1) & \lambda ^2 (\tau -1) \tau  \\
 \sqrt{\tau } \left(\lambda ^2 \tau -1\right) & -\lambda  (\tau -1) \sqrt{\tau } & \lambda ^2 (\tau -1) \tau  & \lambda  (\tau -1) \\
\end{array}
\right),
    \end{equation}
 
\begin{equation}\label{G2}
   G_2= \frac{\sqrt{1-\tau }}{2 \left(1-\lambda ^2 \tau ^2\right)}\left(
\begin{array}{cccc}
 (\lambda  \tau -1) \sin \left(\frac{\phi }{2}\right) & -i (\lambda  \tau +1) \sin \left(\frac{\phi }{2}\right) & (1-\lambda  \tau ) \cos \left(\frac{\phi }{2}\right) & i (\lambda  \tau +1) \cos \left(\frac{\phi }{2}\right) \\
 (1-\lambda  \tau ) \sin \left(\frac{\phi }{2}\right) & -i (\lambda  \tau +1) \sin \left(\frac{\phi }{2}\right) & (\lambda  \tau -1) \cos \left(\frac{\phi }{2}\right) & i (\lambda  \tau +1) \cos \left(\frac{\phi }{2}\right) \\
 \lambda  \sqrt{\tau } (\lambda  \tau -1) \sin \left(\frac{\phi }{2}\right) & i \lambda  \sqrt{\tau } (\lambda  \tau +1) \sin \left(\frac{\phi }{2}\right) & \lambda  \sqrt{\tau } (1-\lambda  \tau ) \cos \left(\frac{\phi }{2}\right) & -i \lambda  \sqrt{\tau } (\lambda  \tau +1) \cos \left(\frac{\phi }{2}\right) \\
 \lambda  \sqrt{\tau } (1-\lambda  \tau ) \sin \left(\frac{\phi }{2}\right) & i \lambda  \sqrt{\tau } (\lambda  \tau +1) \sin \left(\frac{\phi }{2}\right) & \lambda  \sqrt{\tau } (\lambda  \tau -1) \cos \left(\frac{\phi }{2}\right) & -i \lambda  \sqrt{\tau } (\lambda  \tau +1) \cos \left(\frac{\phi }{2}\right) \\
\end{array}
\right),
\end{equation}

    \begin{equation}\label{G3}
     G_3=\frac{1}{4}   \left(
\begin{array}{cccc}
 \dfrac{\lambda  \tau  \cos (\phi )+1}{\lambda  \tau +1} & 0 & \dfrac{\lambda  \tau  \sin (\phi )}{\lambda  \tau +1} & 0 \\
 0 & \dfrac{\lambda  \tau  \cos (\phi )-1}{\lambda  \tau -1} & 0 & \dfrac{\lambda  \tau  \sin (\phi )}{\lambda  \tau -1} \\
 \dfrac{\lambda  \tau  \sin (\phi )}{\lambda  \tau +1} & 0 & \dfrac{1-\lambda  \tau  \cos (\phi )}{\lambda  \tau +1} & 0 \\
 0 & \dfrac{\lambda  \tau  \sin (\phi )}{\lambda  \tau -1} & 0 & \dfrac{\lambda  \tau  \cos (\phi )+1}{1-\lambda  \tau } \\
\end{array}
\right),
    \end{equation}
    and
    \begin{equation}\label{G4}
    G_4= \left( d_x \cos \left(\frac{\phi }{2}\right),d_p \cos \left(\frac{\phi }{2}\right),d_x \sin \left(\frac{\phi }{2}\right),d_p \sin \left(\frac{\phi }{2}\right) \right)^T.
    \end{equation}

\section{Matrices corresponding to Parity detection}\label{Appendix C}
Here, we provide the expressions of the matrices, $Q_1$ and $Q_2$ which appear in the average of the parity operator, (\ref{apari1}) of the main text,

\begin{equation}\label{Q1}
   Q_1=\frac{1}{4}\frac{ 1-\tau}{\lambda ^2 \tau ^2 \cos ^2(\phi )-1} \left(
\begin{array}{cccc}
 \lambda  \tau  \cos ^2(\phi ) & -\cos (\phi ) & \lambda  \sqrt{\tau } \cos (\phi ) & t_0 \\
 -\cos (\phi ) & \lambda  \tau  \cos ^2(\phi ) & t_0 & \lambda  \sqrt{\tau } \cos (\phi ) \\
 \lambda  \sqrt{\tau } \cos (\phi ) & t_0 & \lambda  & -\lambda ^2 \tau  \cos (\phi ) \\
 t_0 & \lambda  \sqrt{\tau } \cos (\phi ) & -\lambda ^2 \tau  \cos (\phi ) & \lambda  \\
\end{array}
\right),
\end{equation}
 with $t_0=\dfrac{\sqrt{\tau } \left(\lambda ^2 \tau  \cos (2 \phi )+\lambda ^2 \tau -2\right)}{2 (\tau -1)}$
 and   \begin{equation}\label{Q2}
     Q_2=\frac{\sqrt{1-\tau }}{2-2 \lambda  \tau  \cos (\phi )}   \left(
\begin{array}{c}
 -\sin (\phi ) \\
 \sin (\phi ) \\
 -\lambda  \sqrt{\tau } \sin (\phi ) \\
 \lambda  \sqrt{\tau } \sin (\phi ) \\
\end{array}
\right).
    \end{equation}
\end{widetext}

%\bibliography{references}
 
 %apsrev4-2.bst 2019-01-14 (MD) hand-edited version of apsrev4-1.bst
%Control: key (0)
%Control: author (8) initials jnrlst
%Control: editor formatted (1) identically to author
%Control: production of article title (0) allowed
%Control: page (0) single
%Control: year (1) truncated
%Control: production of eprint (0) enabled
%

\end{document}